\documentclass[aps,pra,floats,12pt]{revtex4}
\usepackage{graphicx}
\usepackage{amsmath}
\usepackage{amssymb}
\usepackage{color}
\usepackage{ulem}

\newcommand{\w}{{\omega}}

\renewcommand{\l}{{\lambda}}
\newcommand{\G}{{\Gamma}}

\newcommand{\W}{{\Omega}}
\newcommand{\D}{{\Delta}}
\renewcommand{\d}{{\delta}}

\newcommand{\beq}{\begin{equation}}
\newcommand{\eeq}{\end{equation}}
\newcommand{\bea}{\begin{eqnarray}}
\newcommand{\eea}{\end{eqnarray}}

\begin{document}
\title{Interference spectroscopy with coherent anti-Stokes Raman scattering of noisy broadband pulses.}

\begin{abstract} We propose a new technique
for comparing two Raman active samples. The method employs optical
interference of the signals generated via coherent anti-Stokes
Raman scattering (CARS) of broadband laser pulses with noisy
spectra. It does not require spectrally resolved detection, and no
prior knowledge about either the Raman spectrum of the samples, or
the spectrum of the incident light is needed. We study the
proposed method theoretically, and demonstrate it in a
proof-of-principle experiment on Toluene and ortho-Xylene samples.
\end{abstract}

\date{\today}
\author{Evgeny A.~Shapiro$^1$, Stanislav O. Konorov$^1$,  Valery
Milner$^{2}$,~\\\it Departments of Chemistry$^1$ and Physics$^2$,
The University of British Columbia
\\  2036 Main Mall, Vancouver, BC, Canada V6T 1Z1}

\maketitle

\section{Introduction.}

In the last decade, coherent nonlinear optical spectroscopy with
femtosecond pulses has evolved into a powerful tool for chemical
characterization, detection, and microscopy
\cite{Dantus_Appl_Spec_2009, Motzkus_opt_lett_2010, Xie_Phys_Rev_Lett_1999, Potma_PNAS_2001, Konorov_Phys_RevA_ 2009, FemtoCARS-1,FemtoCARS-2,Scully-CARS}. Ultrashort laser pulses exhibit high peak
intensities, which result in high nonlinear signals, at low
average power below the damage threshold of many systems of
interest. Coherent anti-Stokes Raman scattering (CARS) with
time-resolved (tr) and frequency-resolved (fr) detection has been
recently complemented by a number of methods employing shaped
femtosecond pulses. By adjusting the amplitudes and phases of the
spectral components of a broadband pulse, one is able to enhance
Raman excitation of one molecular species while not exciting
another, thus gaining sensitivity to chemical structure
\cite{FemtoCARS-Silberberg}. Pulse shaping also proved useful in
suppressing the non-resonance background signal which usually
reduces CARS sensitivity \cite{Stas-LorentsianProbe, Xu XJG_Opt_Lett_2008, Postma_Opt_Expr_2008, Silberberg_Phys_Rev_Let_2003}. Pulse
shaping approaches to coherent nonlinear spectroscopy typically
rely on the availability of broadband pulses with smooth
well-characterized spectral and temporal profiles
\cite{MIIPS}.

In many situations, the goal of the spectroscopic analysis is a
quick  ``yes'' or ``no'' answer {to the question of whether the
two samples of interest are similar}. One example is a quality
control task in which the sample in question is compared with a
reference.
Here, it is desirable to avoid experiments which require
time-consuming scanning procedures, e.g. delay scanning in tr-CARS
or frequency scanning in fr-CARS detection. Single-shot techniques
based on femtosecond pulse shaping, although very quick, are often
sensitive to the \textit{a priori} knowledge about the anticipated
spectral response from the sample of interest. Another quick
scan-less approach to molecular analysis known as multiplex CARS
(broadband excitation with narrowband probing) relies on
spectrally resolved detection and analysis of the signal.

In this paper we propose and demonstrate a technique for the direct
spectroscopic comparison of two Raman active samples. In the
following text we refer to them as ``reference'' (``$R$'') and
``sample'' (``$S$''), and imply that the goal of the experiment is
to establish the degree of their similarity. The proposed method
does not require time or frequency scanning, and can be
implemented without technically involved pulse shaping. Spectral
analysis of the signal is not required; rather, an integral power
of a single anti-Stokes beam is detected. The scheme does not rely
on the \textit{a priori} knowledge of Raman spectra of either
the sample of interest or the reference. Moreover, it requires
neither characterization nor control over the spectrum of the
input beams. One can employ probe pulses with unknown random
spectral profiles, as long as the latter exhibit sufficiently
narrow features. One example of such spectrum is that of a
transform-limited pulse sent through a random scatterer.

\section{Method}
\begin{figure}
\centering
\includegraphics[width=0.85\columnwidth]{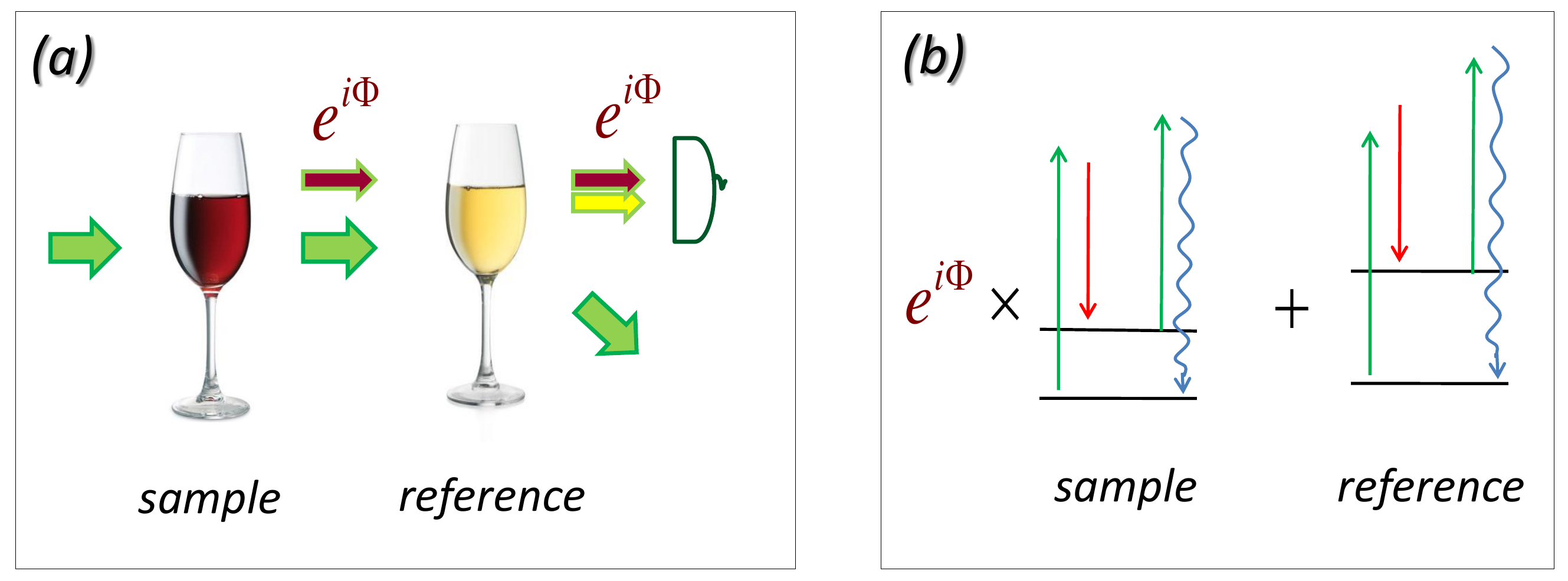} \caption{Color online. The idea of the method.
(a): Two nonlinear signals (thin red and yellow arrows) are generated in two media of interest, Sample and Reference, and interfere at the detector. Variable phase shift $\Phi $ is added to one of the signals.  Interference is strongest when the sample and the reference
are similar. (b): Simplified energy-level diagram for the proposed method. Similar samples generate
ansi-Stokes radiation (curly blue lines) at the same frequency, leading to strong
interference. Chemically different samples generate signal at
different frequencies, leading to no interference. }
\label{Fig-GeneralScheme}
\end{figure}

The idea of the method is shown in Fig.\ref{Fig-GeneralScheme}(a).
Laser pulses pass through the reference and sample - illustrated
in the Figure by two glasses of wine, - generating a nonlinear
spectroscopic response in both of them. If the two materials are
similar (different), their spectroscopic responses are similar
(different) as well. Varying the phase $\Phi$ added to the
nonlinear optical signal between the two media, one gains
100$\%$ interference contrast in the case of identical samples (``$S=R$ ''),
and a lower contrast if $S$ and $R$ differ (``$S\neq R$ ''). As shown below, the
interference contrast is a natural measure of the similarity
between the $S$ and $R$ spectra.

We implement the idea described above using femtosecond CARS
spectroscopy. In femtosecond CARS, broadband pump and Stokes laser
pulses excite a long-lived Raman coherence in the medium. A probe
pulse, which can be either narrow- or broadband, stimulates
radiation at the anti-Stokes frequency
$\w_{AS}=\w_{p}-\w_{S}+\w_{pr}$, where $\w_{p}$, $\w_{S}$,
$\w_{pr}$ are the frequencies of the pump, Stokes, and
probe spectra, respectively. The anti-Stokes signal consists of
two parts. The resonant signal, whose time duration is determined by the life times
of the Raman modes, is given by
\cite{Silberberg-CARS-JCP03,Okamoto-CARS-CP91,KamalovSvirko-CARS92}
\begin{equation}
E_r(\w) =  \int_{-\infty}^{-\infty} d\W  \sum_n \,C_n\,E(\w-\W)
A(\W) \frac{1}{\W-\W_{n}+i\G_{n}} ~.\label{resonanceE}
\end{equation}
Here $n$ enumerates Raman resonances (e.g. vibrational energy
levels of the molecule), $\W_n$ are their energies, $\G_n$ are
their widths,
\begin{equation}
A(\W) = \int_{-\infty}^\infty d\w' E^*(\w'-\W)\,E(\w') \label{Aw}
\end{equation}
{is the two-photon excitation spectrum}, and $E(\w)$ is the
spectral amplitude of the joint pump, Stokes, and probe field. The
non-resonant background (NRB) signal, corresponding to the
instantaneous electronic responce, can be approximated as
\begin{equation}
E_{nr}(\w) = C_{nonres}  \int_{-\infty}^\infty  \,E(\w-\W) A(\W)
\,d\W
 \label{nonresonanceE}
\end{equation}
and can be either weaker or stronger than the resonant one. In
order to separate $E_r$ from $E_{nr}$, one can delay probe pulses
with respect to pump and Stokes pulses, as described below.

\begin{figure}
\centering
\includegraphics[width=0.85\columnwidth]{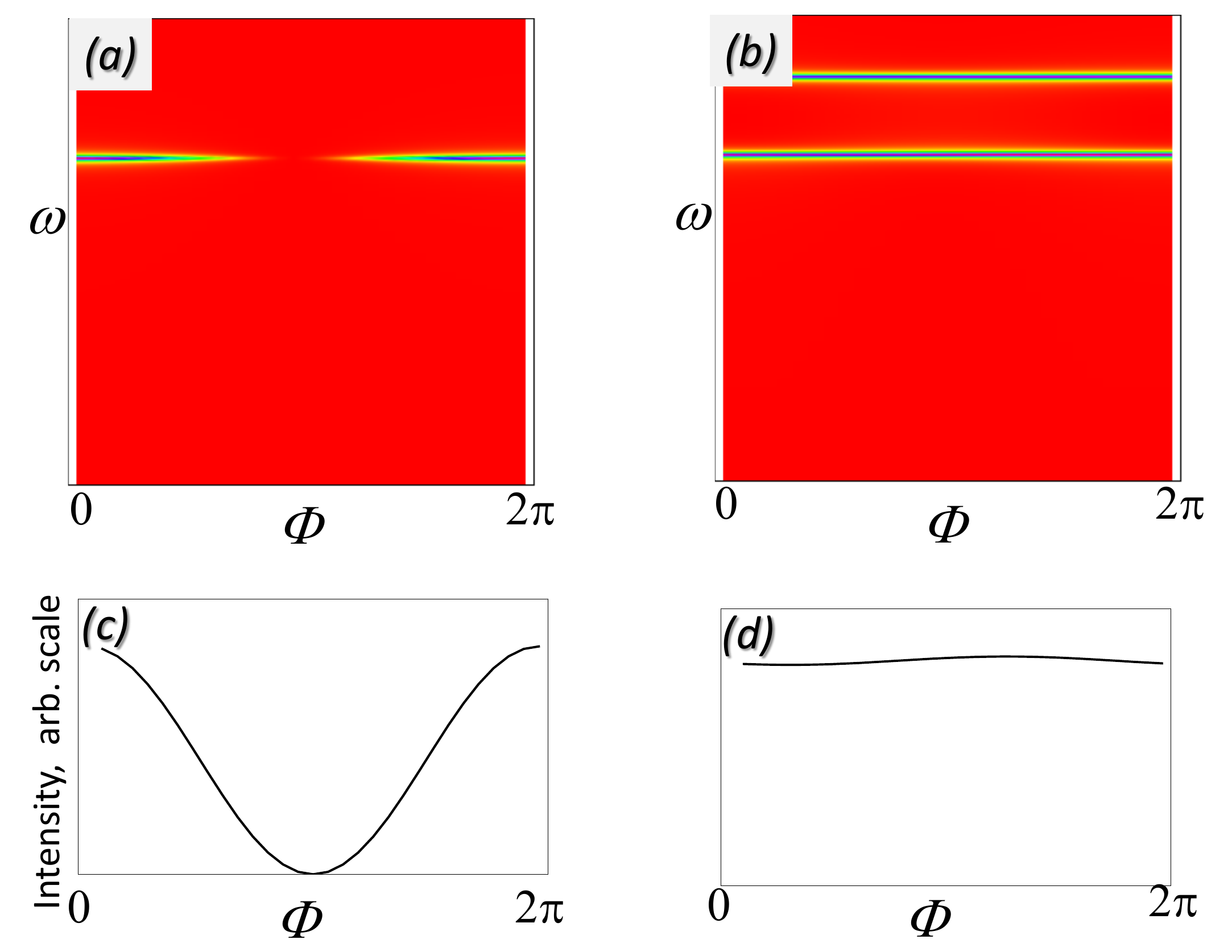}
\caption{Color online. (a,b): Two-dimensional interference maps
for the intensity of multiplex CARS probing two similar (a) and
different (b) samples. (c,d): CARS signals integrated over
frequency. } \label{Fig2-Multiplex}
\end{figure}

\begin{figure}
\centering
\includegraphics[width=0.85\columnwidth]{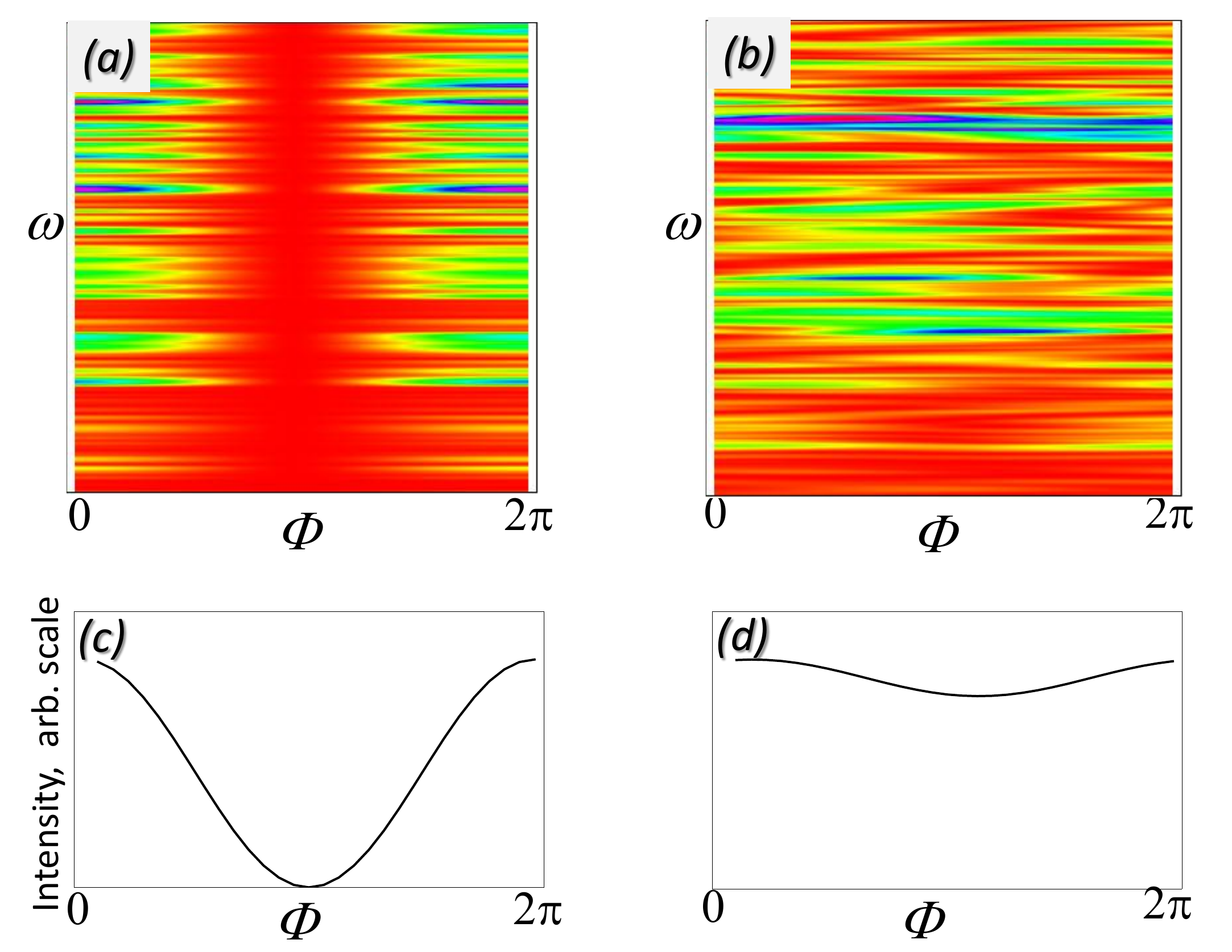}
\caption{Color online. (a,b): Simulated interference map for CARS
with noisy probe with (a): two Toluene samples,  and  (b):
Toluene and ortho-Xylene samples. (c,d): CARS interference
with a noisy probe, averaged over noise realizations and
integrated over frequency.
 (c): two Toluene samples; (d): Toluene and ortho-Xylene samples.} \label{Fig3-Noise}
\end{figure}

Consider, first, a model situation in which both $R$ and $S$ Raman
spectra consist of a single line of width $\G$, centered at $\W_R$
and $\W_S$, respectively (Fig.\ref{Fig-GeneralScheme}(b)). The
idea of our method is most transparent in the multiplex CARS
scheme, where a broadband pump-Stokes excitation is followed by a
narrowband probe. For the latter, we assume that $E_{pr}(\w)=E_0$
within a narrow interval $\d\w$ near the probe frequency
$\w_{pr}$, and $E_{pr}(\w)=0$ otherwise. By our design, the CARS
signal acquires an extra phase factor $e^{i\Phi }$ after exiting
the sample medium. Assuming for simplicity that the broadband
excitation is uniform, i.e. $A(\W)=A_0$ in the relevant frequency
range of the excited Raman modes, we obtain from
Eq.(\ref{resonanceE}) the following resonant signal generated in
both materials:
\begin{equation}
E_r(\w) = \d\w\, E_0 A_0 \left[ \frac{e^{i\Phi}
C_S}{\w-\w_{pr}-\W_S+i\G} + \frac{C_R}{\w-\w_{pr}-\W_R+i\G}\right]
~. \label{Er-multiplex}
\end{equation}
CARS signal from the sample is in the vicinity of
$\w^{S}_{CARS}=\W_S+\w_{pr}$ while for the reference it is near
$\w^{R}_{CARS}=\W_R+\w_{pr}$. When the phase $\Phi $ is scanned,
these signals will interfere if $\W_S = \W_R$, and will not
interfere otherwise (Fig.\ref{Fig2-Multiplex}). To quantify this
interference, we calculate the integral signal by integrating
$E_r(\w)E_r^*(\w)$ over all frequencies. Introducing the
dimensionless frequency $w=(\w-\w_{pr})/\G$, we have from
Eq.(\ref{Er-multiplex})
\begin{equation}
\langle E_r E_r^* \rangle_\w (\Phi) = \d\w^2 |E_0 A|^2\,
\int_{-\infty}^{\infty} dw \,
\left|
\frac{e^{i\,\Phi} C_S}{w-w_S+i} + \frac{C_R}{w-w_R+i}%
 \right|^2
\label{MultiplexTwoResonances-Intensity}
\end{equation}
where $w_{R,S}=\W_{R,S}/\G$, and $\langle\rangle_\w$ stands for
the integration over frequency. The integral in
Eq.(\ref{MultiplexTwoResonances-Intensity}) is readily taken by
contour integration, which yields
%
\begin{equation}
\langle E_r E_r^* \rangle_\w (\Phi) = N_\text{multiplex}  \times
\left[ |C_S|^2 +|C_R|^2 +  \frac{|C_S C_R |}{1+w_{RS}^2}
\cos[\Phi-\D-\phi_C] \right] \label{MultiplexTwoResonances-Final}
\end{equation}
where
\begin{eqnarray}
N_\text{multiplex} &=& 2\pi \d\w^2\,|E_0 A|^2\,       \label{Nmultiplex}  \\
w_{RS} &=& \frac{\W_S-\W_R}{2\G},               \label{MultiplexTwoResonances-Final-wRS}\\
\tan \D &=& w_{RS}                              \label{MultiplexTwoResonances-Final-D}\\
\phi_C &=& \arg [C_S^* C_R ] ~.
                                                  \label{MultiplexTwoResonances-Final-phiC}
\end{eqnarray}
%
Equation (\ref{MultiplexTwoResonances-Final}) is the key for
understanding the proposed method. If $S$ and $R$ are similar,
then $w_{RS} \approx 0$, and the interference contrast in the
integral signal is 100\%: at $\Phi_{min}=\pi$ the resonant CARS
intensity drops to zero, as seen in Fig.\ref{Fig2-Multiplex}(a,c).
If the two samples are completely different, the corresponding
CARS signals are at two different frequencies, and therefore do
not interfere -- this is the situation shown in
Fig.\ref{Fig2-Multiplex}(b,d). Finally, if $S$ and $R$ lines
partially overlap, the integral interference pattern is different
from that of two similar samples in two ways. First, its contrast
is lower. If the $S$ and $R$ signals are of the same strength,
$C_S=C_R$, then the contrast is reduced due the factor
$1/(1+w_{RS}^2)$ in the square brackets of
Eq.(\ref{MultiplexTwoResonances-Final}). From
Eq.(\ref{MultiplexTwoResonances-Final-wRS}), one can see that
$w_{RS}$ may indeed be viewed as a natural measure of the mismatch
between the two Raman spectra. Second, the minimum and maximum of
the integral signal $\langle E_r E_r^* \rangle_\w (\Phi)$ are
offset from their values $\Phi_{min}=\pi$ and $\Phi_{max}=0$ by
the value $\D+\phi_C$.
This situation is illustrated in Figs.\ref{Fig2-Multiplex} and
\ref{Fig3-Noise}.

We now show that having a single narrowband probe pulse (as in
multiplex CARS) is not essential for the proposed interference
detection. Let us illuminate the $S$ and $R$ media with
transform-limited broadband pump and Stokes pulses, and a probe
pulse that consists of $N$ narrowband components with an average
spectral width $\d\w$. The spectral separation between these
components is not important: one may employ a set of well isolated
lines, as well as a single broadband line with a random phase
modulation under its spectral envelope. The only relevant
parameter is the spectral correlation length $\d\w$ of the complex
probe field. It is important, however, that the phases of $N$
probe's spectral components are not correlated with each other.

The resonant anti-Stokes response now has the form
\begin{equation}
E_r(\w) = \d\w\, \sum_{i=1}^N E_i A(\w-\w_{pr\,i})
\left[\frac{e^{i\Phi} C_S}{\w-\w_{pr\,i}-\W_S+i\G} +
\frac{C_R}{\w-\w_{pr\,i}-\W_R+i\G} \right] \label{Er-NoisyProbe}
\end{equation}
This equation is similar to Eq.(\ref{Er-multiplex}), except that
the single term $E_0 A_0$ is replaced by a sum of $N$ complex
amplitudes $E_i= |E_i| \exp[i\phi_i]$ multiplied by the Raman
excitation amplitudes $A(\w-\w_{pr\,i})$. The intensity of the
CARS signal $E_r(\w)E_r^*(\w)$ thus contains $N^2$ terms,
\begin{eqnarray}
E_r(\w)E_r^*(\w) &=& \d\w^2 \, \sum_{i,j} |E_i E_j|
e^{i(\phi_i-\phi_j)}\, A(\w-\w_{pr\,i}) A^*(\w-\w_{pr\,j})
\nonumber \\
&&\times\left[ \frac{e^{i\,\Phi}\,C_S}{\w-\w_{pr\,i}-\W_S+i\G} +
\frac{C_R}{\w-\w_{pr\,i}-\W_R+i\G} %
 \right] %
\nonumber \\ %
&&\times \left[
\frac{e^{-i\,\Phi}\,C_S^*}{\w-\w_{pr\,j}-\W_S-i\G} +%
\frac{C_R^*}{\w-\w_{pr\,j}-\W_R-i\G} %
 \right] ~.
\label{TwoResonances-Intensity-noaveraging}
\end{eqnarray}
Within a given realization of a random probe spectrum, the
interference pattern in the $(\w,\Phi)$ plane can be quite
complex. CARS signal consists of many spectral lines, reflecting
both the $S$ and $R$ Raman resonances convolved with the probe
spectrum. Two examples are shown in Fig.\ref{Fig3-Noise}. If the
$S=R$, the two terms in the square brackets of
Eq.(\ref{Er-NoisyProbe}) are similar, and the interference
contrast at each frequency is high (Fig.\ref{Fig3-Noise}(a)).
However, if the two materials are different, the cross terms in
Eq.(\ref{TwoResonances-Intensity-noaveraging}) contribute to the
interference pattern in a random uncorrelated way. Depending on
the spectral amplitudes $|E_i|$ and phase differences
$\phi_i-\phi_j$, the interference fringes at different frequencies
lose their contrast and cease to occur at the same values of
$\Phi$ . As a result, the intensity map in the $(\w,\Phi)$ plane
shows no regular fringe structure, as demonstrated in
Fig.\ref{Fig3-Noise}(b).


The apparent randomness of the CARS intensity map disappears once
it is averaged over several noise realizations. Since the phases
$\phi_i$ of different spectral components of the probe field are
not correlated, the cross terms in
Eq.(\ref{TwoResonances-Intensity-noaveraging}) average out, while
those with $i=j$ do not. The resulting expression is similar to
Eq.(\ref{MultiplexTwoResonances-Intensity}), but with a
normalization factor accounting for having $N$ spectral lines in
the probe. By repeating the steps which lead to Equation
(\ref{MultiplexTwoResonances-Final}), we obtain, for the case of a
noisy probe pulse, an expression of exactly the same form. The
only difference is the normalization constant $N_\text{noise}$ :
\begin{equation}
  N_\text{noise} = 2\pi \d\w^2 |A_0|^2 \sum_{i=1}^{N}\langle|E_{pr,\,i}|^2\rangle        \label{Nnoise}
\end{equation}
where $\langle .. \rangle$ denotes averaging over noise
realizations, and $A_0$ is the Raman excitation probability, which
is assumed to be uniform in the relevant frequency range. If all
the spectral lines of the probe pulse are of the same intensity,
the integrated CARS intensity is simply $N$ times that generated
by a single spectral component.

If broadband pump and Stokes pulses excite several vibrational
resonances, one has to add in
Eq.(\ref{MultiplexTwoResonances-Final}) a sum over all $S$ and $R$
lines. For different substances, the lines will, in general, be
randomly placed with respect to each other. The value $w_{RS}$
(Eq.\ref{MultiplexTwoResonances-Final-wRS}) will be positive for
some pairs of lines and negative for others. After summation over
$\Omega _{S}$ and $\Omega _{R}$, the minima and maxima of the
resulting interference pattern may not shift much from
$\Phi_{min}=\pi$ and $\Phi_{max}=0$. Yet the contrast of the
interference pattern will remain low if the $S$ and $R$ materials
differ significantly. The contrast depends on the average value of
$w_{RS}$ -- a natural measure of the similarity of the two
spectra.


We numerically simulated an interference of two noisy CARS signals
from Toluene (C$_7$H$_8$) and o-Xylene (C$_8$H$_{10}$), whose
Raman spectra are well known. The excitation pulse parameters were
similar to those used in our experiments and described in
Sec.\ref{Sec-Experiment}. Fig.\ref{Fig3-Noise}(a,b) shows two
calculated interference patterns. In plot (a), Toluene was used as
both the Sample and Reference, whereas replacing the Sample by
o-Xylene resulted in plot (b). Integration over frequency produced
the results which are plotted in panels (c,d) of the Figure and
demonstrate a striking difference in the interference contrast for
the cases of Toluene--Toluene and Toluene--o-Xylene interference.
Within a wide set of parameters, the simulations showed a 100\%
fringe visibility for the interference of two identical
substances, and only  $\sim$17\% visibility for a
Toluene--o-Xylene combination, given an equal strength of their
resonant signals. The $\sim$17\% visibility of the
Toluene--o-Xylene interference is due to the partial overlap of
their Raman spectral lines, and thus gives the measure of their
spectral similarity in accord with
Eq.(\ref{MultiplexTwoResonances-Final}).

Our numerical tests allowed us to make several observations.
First, the integrated picture is insensitive to the spectral shapes
of pump, Stokes, and probe pulses as long as  the correlation
length of noise introduced to the probe spectrum is smaller than
the average separation between the $S$ and $R$ Raman lines.
Second, having many independent lines in the probe spectrum is
advantageous as compared to having a few lines. In the case of many
lines, not only the signal is stronger (see Eq.(\ref{Nnoise})),
but one also needs to average the interference map over fewer
noise realizations. Finally, we noticed that using random pulses for the
{\it excitation}, as well as for probing, leads to qualitatively
similar results, although requires averaging over large numbers
of noise realizations, and gives higher interference contrasts
for the ``$S\neq R$'' case. Assuming non-equal signal
strengths further reduced the interference visibility, enabling an
easy distinction between the similar-samples and different-samples
combinations.

As expected and confirmed by our simulations, non-resonant
background increases the fringe visibility even if the interfering
signals are generated by two different materials. Indeed, NRB from
Sample is identical to that from Reference, and their
interference produces high-contrast oscillations with $\Phi $.
Hence, as in all CARS schemes, the proposed method requires
suppressing the non-resonant signal as much as possible. This task
is difficult to accomplish in multiplex CARS, where the overlap of
all three excitation pulses in time is large. Ideally, one needs a
long probe pulse of the duration comparable to the lifetime of the
Raman coherence, but with a short front edge delayed in time with
respect to the pump-Stokes excitation pulses
\cite{Stas-LorentsianProbe}. Such an optical field can be
generated by propagating an ultrashort pulse through a randomly
scattering medium: the sharp front edge corresponds to the
ballistic part, followed by a long tail of randomly scattered
photons \cite{Lagendijk-Johnson03}.

In the final series of calculations, instead of introducing random
phase modulation in the probe spectrum, we used the transmission
spectrum of a random layered medium, calculated using the transfer
matrix method \cite{Pendry}. Our simulations show that the
integral interference curves of Fig.\ref{Fig3-Noise} are well
reproduced as long as the widths of well separated un-correlated
lines in the probe spectrum are smaller than the differences
between the Raman lines. We refer the reader to
Ref.\cite{nascars3} for a detailed discussion of the transmission
lines of a random layered medium, and of their dependance on the
system parameters.

\section{\label{Sec-Experiment}Experimental procedure and results} 
\begin{figure}[b]
\centering
\includegraphics[width=0.85\columnwidth]{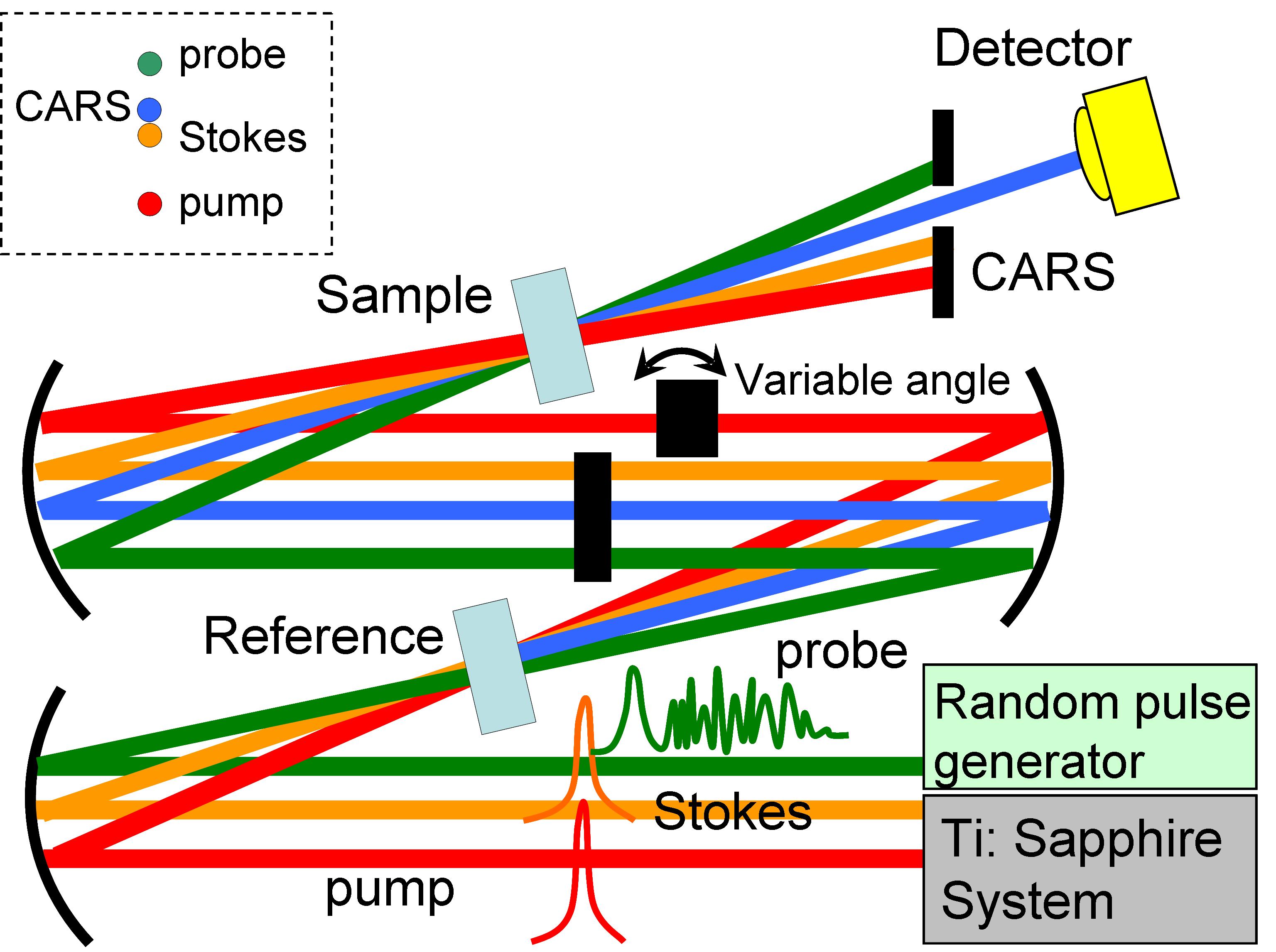}
\caption{Experimental setup. Femtosecond Ti:Sapphire laser system
with OPA generates pump pulses at 1240 nm (signal) and Stokes
pulses at 1125 nm (second harmonic of idler). Probe pulses at 800
nm were coupled into a spectral pulse shaper. The shaper produced
probe pulses with a random spectral profile. All three beams were
collimated in a vertical plane (insert) and focused into the first
(Reference) cuvette. The beams were then collimated. Probe, Stokes
and CARS beams passed through a fixed coverslip glass. Pump
pulses, on the other hand, passed trough a similar coverslip glass
mounted on a rotation stage. All beams were finally focused into
the second (Sample) cuvette. CARS signal was spatially separated
and coupled into a spectrometer.}
  \vskip -.1truein
\label{Fig-ExpSetup}
\end{figure}

Experimental results have been obtained using
the setup shown in Fig.\ref{Fig-ExpSetup}. It consisted of a laser
system based on a femtosecond Ti:Sapphire oscillator (Synergy,
Femtolasers), a regenerative amplifier (Spitfire Pro, Spectra
Physics) and an optical parametric amplifier (OPA) (Topas, Light
Conversion). The amplifier generated 3 mJ, 35 fs pulses at the
central wavelength of 800 nm and 1 kHz repetition rate. A portion
of the 800 nm beam was coupled into a home-built spectral shaper
based on a 640-element liquid crystal spatial light modulator
(CRI, USA). The shaper produced probe pulses with a random
spectral profile. The latter consisted of 25 randomly distributed
Lorentzians of 1 nm width (full width at half maximum) and random
relative phases, as required in the proposed approach. One example
of such a random probe spectrum is shown in Fig.\ref{Fig-Shaper
spectra}(a). The corresponding temporal envelope of the pulse is
plotted in panel (b), which demonstrates another key feature of
the used pulse shape -- a sharp rising edge and a long random
pulse train behind it.

\begin{figure}[b]
\centering
\includegraphics[width=0.6\columnwidth]{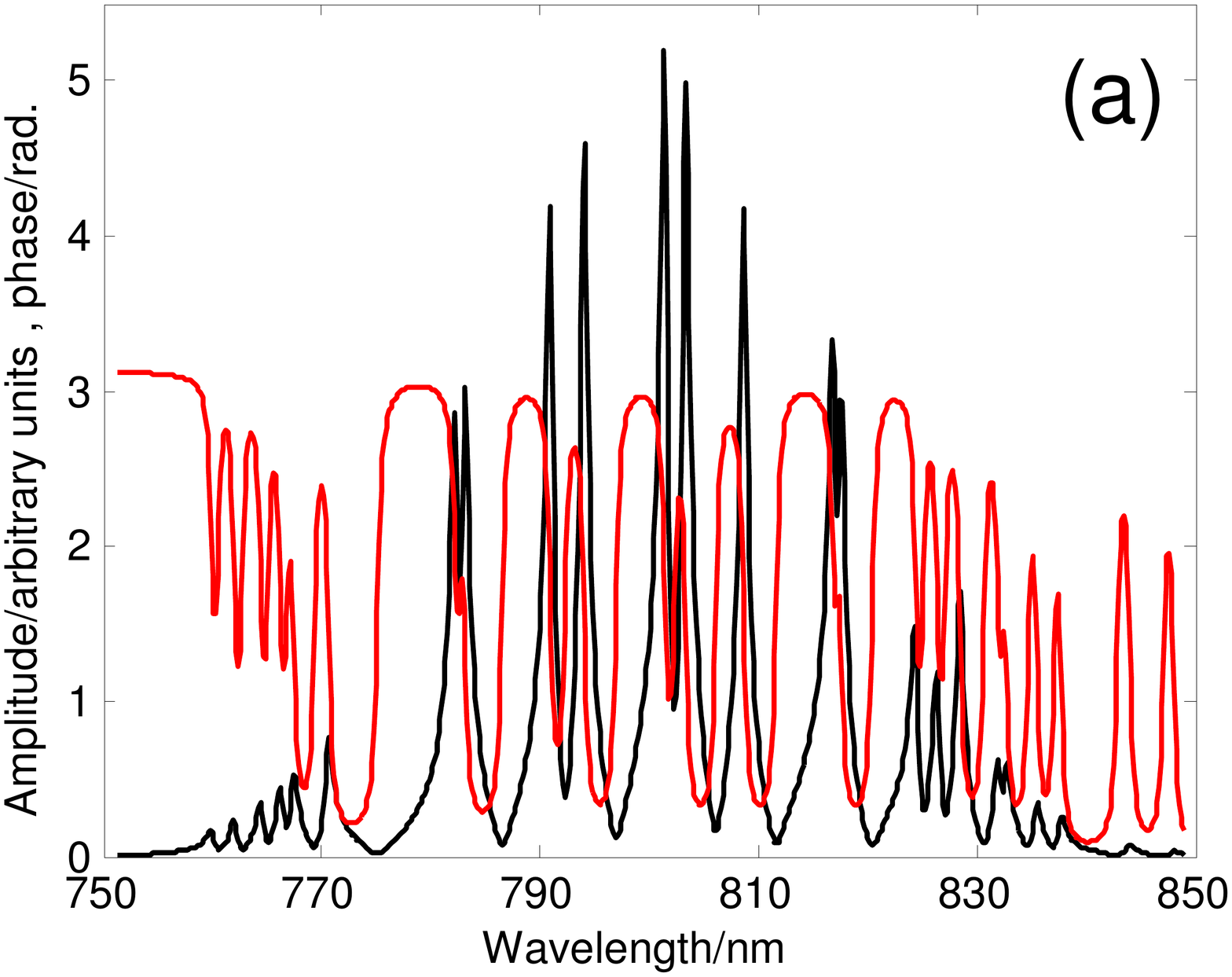}
\includegraphics[width=0.6\columnwidth]{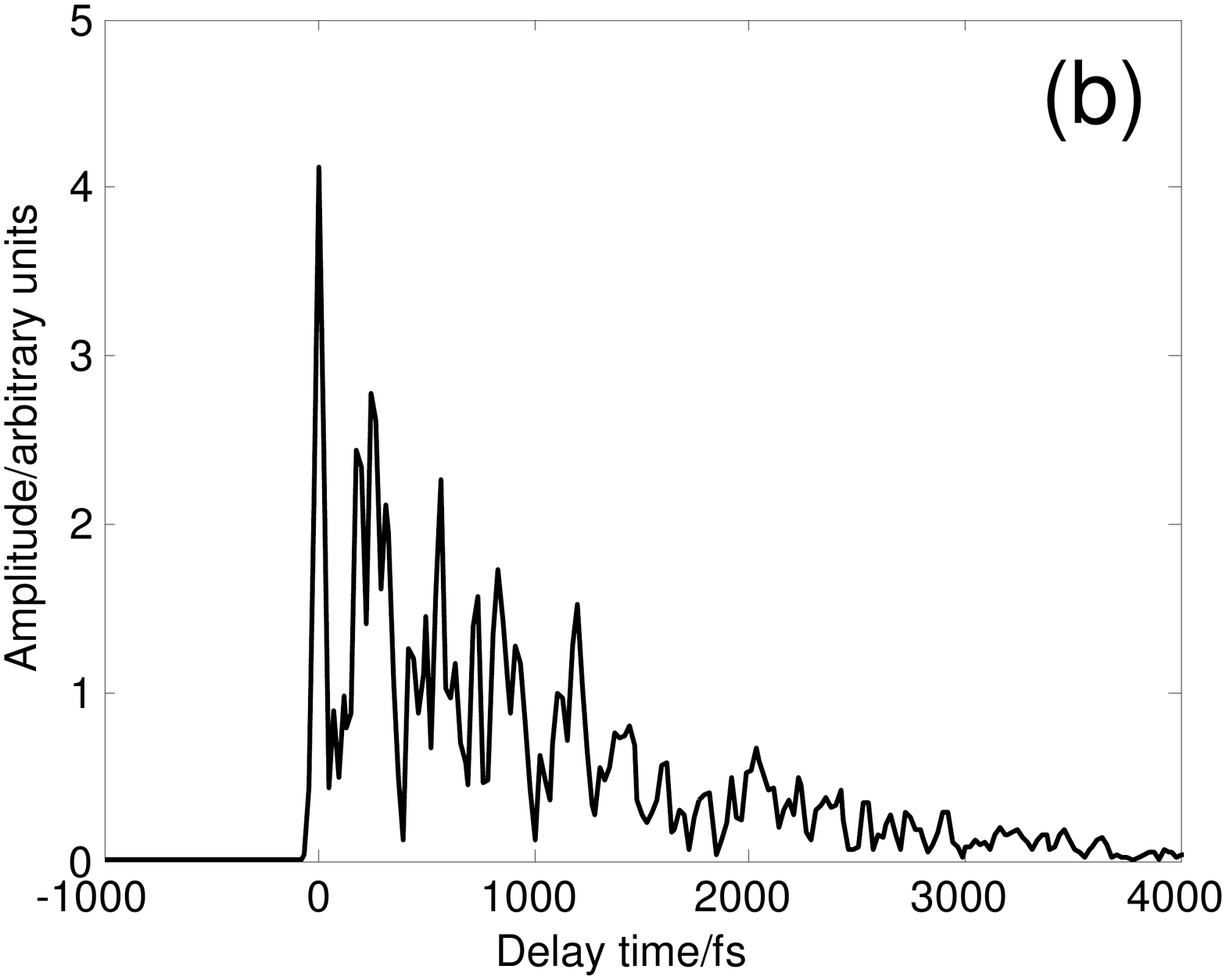}
\caption{An example of the probe pulse used in the experiment. a)
Spectral intensity (solid black) and phase (dotted red); b)
Temporal profile.}
  \vskip -.1truein
\label{Fig-Shaper spectra}
\end{figure}

Another part (1 mJ) of the 800 nm beam was used to pump an OPA,
which produced Stokes and pump pulses at 1240 (signal) and 1125 nm
(second harmonic of idler), respectively. These pulses were
synchronized in time, whereas the front edge of probe pulses was
delayed by 200 fs with respect to the overlapping pump-Stokes
pairs. All three beams were collimated in a vertical plane (see
insert in Fig.\ref{Fig-ExpSetup}) and focused with a 25 cm focal
distance silver mirror into a 200 $\mu$m optical path cuvette with
50 $\mu$m thick walls serving as Reference.

The beams were then collimated by the second 25 cm focal distance silver mirror. Probe and Stokes beams, together with the generated CARS beam passed through a fixed 50 $\mu$m coverslip glass. Pump pulses, on the other hand, passed trough a similar 50 $\mu$m coverslip glass mounted on a rotation stage. The latter provided a variable phase shift for producing the interference fringes as required by our method. All beams were finally focused into the second cuvette with another 25 cm focal distance silver mirror. CARS signal was spatially separated and coupled into a spectrometer (Model 2035, McPherson) operating with the spectral resolution of 0.5 nm and equipped with a cooled CCD camera (iDus, Andor). Exposure time was set to 0.5 seconds, and CARS spectrum was recorded as a function of the coverslip angle $\theta $. The energy of all input beams was set at 3 $\mu$J per pulse. Reference cuvette was filled with Toluene, whereas Sample cuvette contained either Toluene or o-Xylene.

\begin{figure}
\centering
\includegraphics[width=0.8\columnwidth]{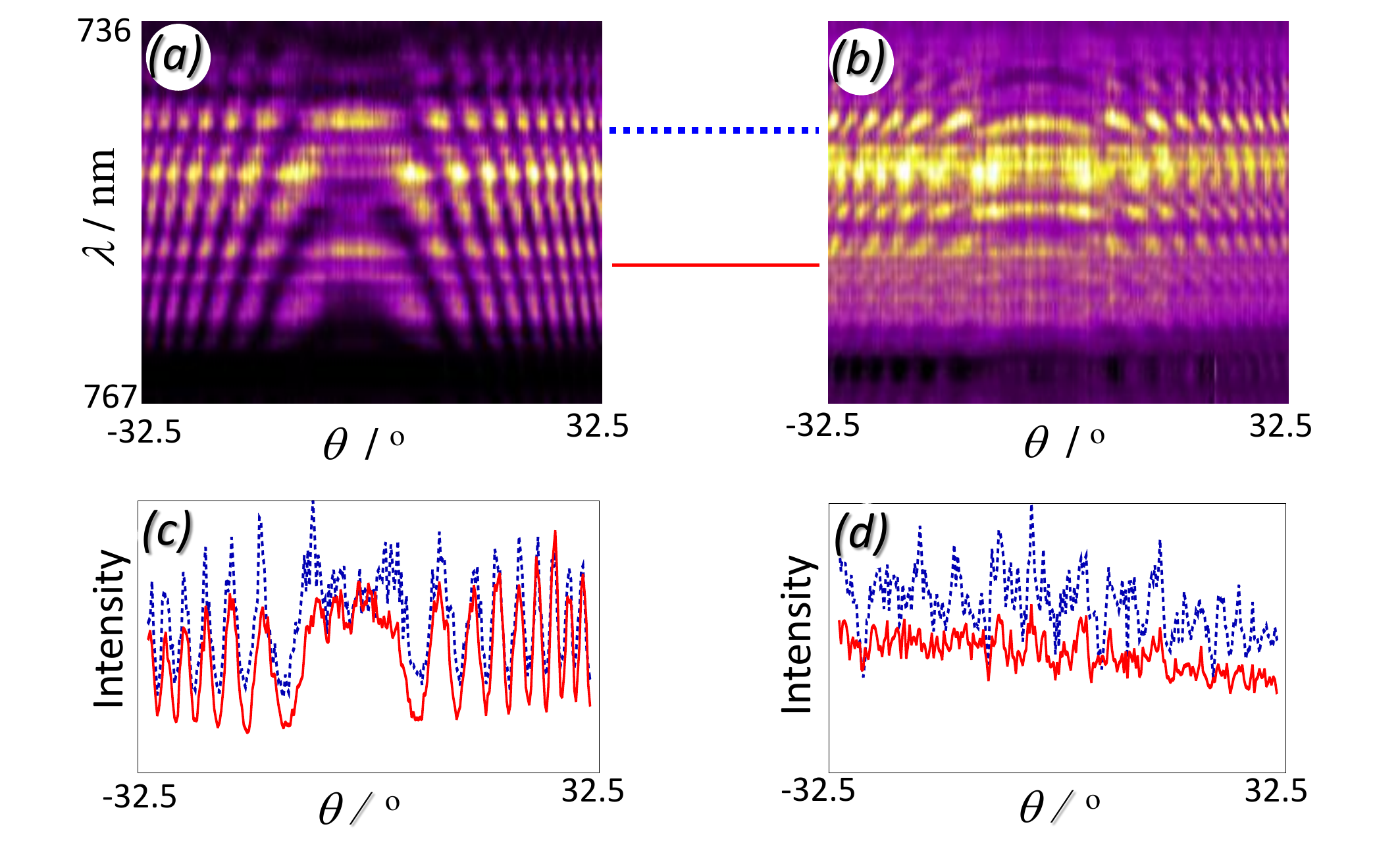}
\caption{Color online. Experimental results. Two-dimensional
interference patterns for Tolune-Toluene (a) and Toluene-o-Xylene
combinations (b). (c,d): Respective cross-sections of the 2D maps at two fixed wavelengths denoted by dashed blue and solid red lines.
values. } \label{Fig6-Exp}
\end{figure}

 Fig.\ref{Fig6-Exp}(a,b) shows the
recorded CARS interference signal for the Toluene-Toluene and
Toluene-o-Xylene combinations. The pattern was averaged over five
realizations of the random spectrum of probe pulses. High
visibility interference fringes are clearly seen in plot (a),
confirming that the two materials -- Reference and Sample, are
very similar in their CARS response. Unfortunately, the parabolic
shape of the fringes in the $(\Phi,\l)$ plane did not allow us to
integrate the signal over frequency while preserving the fringe
contrast. We attribute the fringe curvature to chromatic
dispersion, which causes an additional phase accumulation between
the $S$ and $R$ media, and cannot be easily compensated. Similar
pattern was observed in the interference pattern for the
non-resonant signals, and reproduced in our numerical analysis.
We note that in the absence of dispersion, our calculations
predict that the 2D interference map for the case of ``$S=R$'' is a
series of vertical strips. Hence no frequency-resolved detection will be necessary to detect the fringe contrast in the absence of dispersion.

Even though integrating the two-dimensional experimental
signal over $\omega $, and hence eliminating the need for a
spectrometer, was not possible at this stage, we demonstrate that
the fringe visibility at each wavelength is quite strong. This is
seen in Fig.\ref{Fig6-Exp}(c), which shows two cross sections of
the full map taken at two arbitrary values of $\l$ (two horizontal
lines between plots (a) and (b)). In contrast, both the 2D
interference map and the 1D cross sections for the
Toluene--o-Xylene combination, shown respectively in
Figs.\ref{Fig6-Exp}(b) and (d), exhibit irregular fringes with
 much lower visibility. This qualitative result of the
proof-of-principle experiment confirms the feasibility of the
proposed method.

\section{Summary and discussion}
We have shown theoretically, and demonstrated experimentally, that the interference of two CARS signals obtained by scattering noisy light from two Raman active samples offers a good measure of their chemical similarity. Provided the concentrations are similar, the interference is determined by a single parameter $w_{RS}$ (Eq.(\ref{MultiplexTwoResonances-Final-wRS})), which represents the degree of overlap between the Raman lines, averaged over all pairs of lines. Even though spectrally resolved detection was required at the present stage of the experimental development, we plan to eliminate this requirement in the future work. This could be done by either compensating the residual chromatic dispersion, or by filtering the detected signal with a band-pass filter, which was effectively demonstrated in this work.

For accommodating the samples of significantly different concentration, our method will have to include an additional calibration step.  Using the non-resonant background, the $R$ and $S$ contributions are first equalized by attenuating one of the two CARS signals until the contrast of the non-resonant interference reaches 100\%. After that, NRB is eliminated by delaying probe pulses and the resonant interference signal is detected. Further, excitation in the fingerprint region, rather than across the whole Raman active range, can be used for increasing the detection sensitivity. Finally, the required noisy probe can be generated by sending a broadband pulse through a randomly scattering medium as discussed in the text above. With these improvements, the proposed interference method may become a powerful tool for a quick preliminary test: if the interference contrast is above a certain threshold, a more accurate frequency- or time-resolved analysis is executed.

By exploiting the interference of two noisy pulses for retrieving the information about the samples in question, our method employs a general principle known as Coherence-Observation-by-Interference-Noise (``COIN'') \cite{Prior-COIN-PRL95,COINexp-JCP01,Leichtle-COIN-98}. In COIN, the interference of two noisy signals is used to deduce the degree of their coherence. In our case, the latter is equivalent to the degree of their spectral similarity. If the two spectra of interest are similar, the two signals (however noisy!) will interfere and produce fringes of high contrast. When, on the other hand, the spectra are different, the interference is suppressed and the the fringe visibility decreases. Random spectral noise has been recently exploited in coherent nonlinear spectroscopy \cite{nascars,nascars2,nascars3}, and this work extends its applications to interferometric CARS techniques.

This work was supported by DTRA, CFI and NSERC.

\end{document}